\documentclass[letterpaper]{jpconf}
\usepackage{graphicx}

\newcommand{\pbarp} {$\bar{p}/p~$}
\newcommand{\snn} {$\sqrt{s_{NN}}$}
\newcommand{\srhic} {$ \sqrt{s_{NN}}= {\rm 200  GeV}$}
\newcommand{\sqrts}[1]            {$\sqrt{s} = #1~{\rm GeV}$}
\newcommand{\pT}               {$p_{\rm{T}}$}

\begin{document}

\title{Stopping and Baryon Transport in Heavy Ion Reactions}
\author{F.Videb\ae k$^1$}
\address{$^1$ Physics Department,
Brookhaven National Laboratory,
Upton, NY11973, USA}
\ead{videbaek@bnl.gov}

\begin{abstract}
In this report I will give an experimental overview on nuclear stopping in hadron collisions, 
and relate observations to understanding of baryon transport. 
Baryon number transport is not only evidenced via net-proton distributions 
but also by the enhancement of strange baryons near mid-rapidity. 
Although the focus is on high-energy data obtained from pp 
and heavy ions from RHIC, relevant data from SPS and ISR will be considered. 
A discussion how the available data at higher energy relates 
and gives information on baryon junction, quark-diquark breaking will be made.
\end{abstract}
%
%
\section{Introduction}

A major goal for relativistic heavy ion reactions is to form hot nuclear matter at energy densities in clear excess 
over the value ($\approx 1 \rm{GeV/fm^{3}}$) predicted from Lattice QCD needed to make the phase transition or cross 
over to matter dominated by degrees of freedom  of quarks and gluons rather than of hadrons. 
The next goal is to study and quantify the properties to confront those with properties of non-perturbative QCD. 
Such experiments has been carried out at RHIC with considerable progress and remarkable results from the
first 3 years of experiments. Please refer to the white-papers from the 4 experiments in Refs.\cite{whitepapers}.
The deposited energy is essential to understand the formation of this medium. 
Because baryon number is conserved, and rapidity distributions are only slightly affected by 
rescattering in the later stages of the collisions, the measured baryon distribution retains 
information about energy loss and allows the degree of nuclear stopping to be determined. 
Such measurements may also help distinguish between different mechanism for transporting
baryons to mid-rapidity. 

At very low energies $1-15$~AGeV the hadrons preserve their identity with multiple collisions and excitation 
to resonances being important ingredient in the description of both stopping, transverse momentum spectra 
as well as strange particle production. 
At the higher energies partonic degrees become important, and many features can be described
using a string picture. 
At these higher energies it have long be thought that the dominant mechanism for transport of baryon number is 
that of quark-diquark breaking (of the strings) where the baryon
number is carried(associated) with the valence quarks. I.e. the distribution will reflect the distribution $q(x)-\bar{q}(x)$.
Such a mechanism is not able to move the net-baryon number over a large range of $x$ .
These distribution are flat in $x$~( $e^{-y}$ in rapidity ) for a single collision as observed at SPS energies.
Already an analysis of the ISR data in pp collisions \cite{Kharzeev:BaryonJunction,Kopeliovich:1998za}, 
and later data from HERA that shows a non-zero baryon asymmetry of $\approx 8\%$ in $\gamma p$ reaction 
at more than 7 units of rapidity from the incident baryon~\cite{Kopeliovich:1999bv} has demonstrated that additional mechanisms
 with a slower $x$ and rapidity dependence are needed to describe the data. One such mechanism 
described in the afore mentioned publications is the baryon junction originally proposed in Ref.~\cite{Rossi:1977cy}.
The baryon junction can be thought of as a final state where the incident quarks couple to a color decuplet
state, or a topological structure where three gluons join in a junction carrying the 
baryon number with the valence quarks left in high rapidity mesons.
This allows the baryon number to be carried to a much lower value of $x$ through a diminished $x$ (or \snn) dependence of
the cross section. So even if the probability for such mechanism is small it may become important 
and possibly dominant at higher energies.  It should be mentioned that other mechanism has been proposed 
in Ref.~\cite{bass:2003baryons} (parton cascade) and Ref.\cite{CapellaKopeliovich:1996} (diquark breaking) 
that may also be relevant for baryon transport. 
Clearly experimental data from both pp and AA are needed if we are to distinguish between these possibilities.
In this paper I will review data in pp and AA collisions that sheds light on the issue of baryon transport and will compare
data to predictions of  models with and without the mechanism of baryon junction included. 
Some basic ideas  comes from Refs.\cite{FVOH,BRAHMS:2004netproton}. See a recent paper\cite{ToporPop:2004}
which has discussion as well as a wealth of references to models relevant to baryon dynamics.

  \begin{figure}[!ht]
\begin{center}
\resizebox{0.55\textwidth}{!}
           {\includegraphics{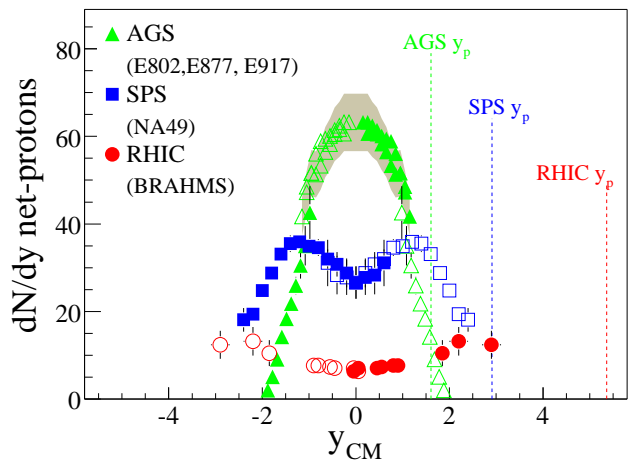}}
\end{center}
\caption{\label{fig:netprot_energy}Rapidity density of net protons (i.e.\
      number of protons minus number of antiprotons) measured at AGS, SPS, and
      RHIC for central collisions.}
\end{figure}

\section{Experimental Considerations}

The stopping in nuclear collisions can be estimated from the rapidity loss experienced
by the baryons in the colliding nuclei. If incoming baryons
have rapidity, $y_b$ in the C.M. system and the average rapidity
\begin{equation}
<y> = \int_0^{y_b} y {{dN}\over{dy}} dy  / \int_0^{y_b}
{{dN}\over{dy}} dy
\end{equation}
after the collision, the average rapidity loss is $\delta y = y_b
- $$<$$y$$>$~\cite{FVOH,BRAHMS:2004netproton}.
Here $dN/dy$ denotes the
number of net-baryons (number of baryons minus number of
antibaryons) per unit of rapidity. In the case of full
stopping $\delta y$ approaches $ y_b$. 
Thus, the distributions of $dN/dy$ should be
known from mid-rapidity to beam rapidity. 
Usually the measurements are for protons, in some cases for
$\Lambda s$ while rarely have the neutrons been measured. To get the net baryon distributions corrections 
and extrapolations have to be made. 
At SIS energies rather detailed measurements have been obtained at 0.4~AGeV and 1.5~AGeV \cite{FOPI:2003bh}.
The clever use of medium mass beams with different isospin content has shown that already at 1.5~AGeV the 
system is not fully stopped, but has a small degree of transparency.

\begin{figure}[!h]
\begin{center}
\resizebox{0.55\textwidth}{!}
           {\includegraphics{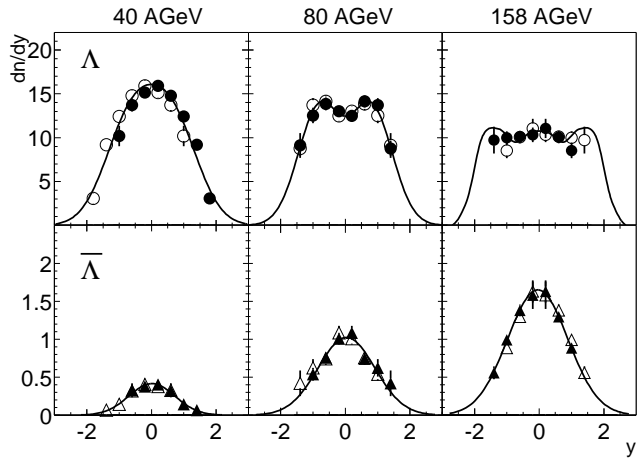}}
\end{center}
\caption{\label{fig:na49lambda}
Rapidity distributions of $\Lambda$ and ${\bar{\Lambda}}$ at 40, 80 , and 158 AGeV beam energy. 
The figure is from Ref.\cite{NA49:lambda}.
}
\end{figure}

At AGS energies the number of produced antiprotons is very small
and the net-baryon distribution is similar to the proton
distribution~\cite{chap3ref3,chap3ref4,chap3ref5}. The net-proton
rapidity distribution is centered around $y=0$ and is rather
narrow (Fig.\ref{fig:netprot_energy}
 The rapidity loss is about 1 for a beam rapidity of
$\approx 1.6$ . At CERN-SPS energies ($\sqrt{s_{NN}}= 17~{GeV},
158 {AGeV}$ Pb+ Pb reactions) the rapidity loss is 1.75
for a beam rapidity of 2.9~\cite{NA49:netproton}, about the
same relative rapidity loss as at the AGS.

At SPS another feature
is visible (see Fig. \ref{fig:netprot_energy}). 
The net proton rapidity distribution
shows a double 'hump' with a dip around $y=0$. This shape results
from the finite rapidity loss of the colliding nuclei and the
finite width of each of the humps, which reflect the rapidity
distributions of the protons after the collisions. This picture
suggests that the reaction at the SPS is beginning to be
transparent in the sense that fewer of the original baryons are
found at midrapidity after the collisions, in contrast to the
situation at lower energies. 
The net-$\Lambda$ distributions on the other hand do not show this bump. The data
from NA49 \cite{NA49:lambda} displayed in Fig.\ref{fig:na49lambda} show a rather flat distribution. 
This is certainly in part due to the higher inelasticity required to produced strange baryons.
Also these data shows that at SPS the hyperon production is significant and must be considered
for the net baryon distributions. A ratios of $\Lambda \over{p}$ of $\sim 0.4$ is observed at SPS.

BRAHMS has measured the net proton rapidity distribution at RHIC
in the interval $y=0-3$ for  central  ($0-10\%$)
Au+Au collisions at \srhic. Details of the analysis can be found in
\cite{BRAHMS:2004netproton}. The results are displayed in
Fig.\ref{fig:netprot_energy}. 
The distribution measured at RHIC is both qualitatively and quantitatively very different
from those at lower energies indicating a significantly different
system is formed near mid-rapidity.
At RHIC the $\Lambda$ production is even more crucial for estimation of the net-baryon yield as well as correct for in the
measured proton and anti-proton yields. 
A detailed discussion can be found in \cite{BRAHMS:2004netproton}. 
The ratio $\Lambda\over{p}$ of $ \sim 0.9$ was observed by STAR and PHENIX 
near midrapidity\cite{star:2002lambda,phenix:2002lambda}.
The BRAHMS analysis presented assumes that the measured  ratios at mid-rapidity
is representative for all measured rapidities up to 3.

\begin{figure}[!ht]
\begin{center}
\resizebox{0.55\textwidth}{!}
           {\includegraphics{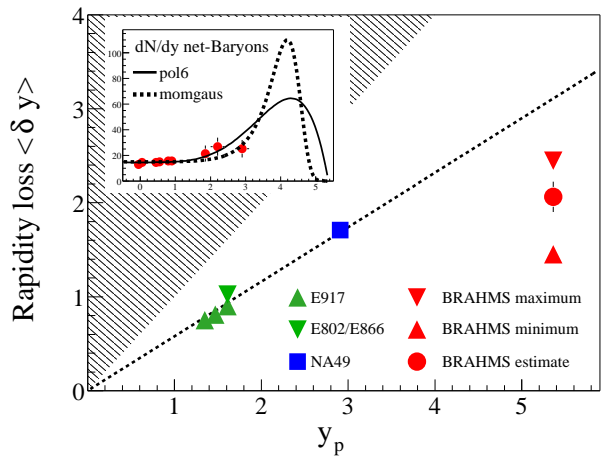}}
\end{center}
 \caption{\label{fig:dyvs_energy}Average rapidity loss as deduced from net-proton distributions vs.
beam rapidity. The straight line is the linear extrapolation from Ref.\cite{FVOH} for constant relative rapidity loss.
Insert: two possible net-baryon distributions (Gaussian in $p_T$ and 6'th
      order polynomial) respecting baryon number conservation. 
}
\end{figure}

The net number of protons per unit of rapidity around $y=0$ is
only about 7 and the distribution is flat over at least the $\pm
1$ unit of rapidity. The distribution rises in the rapidity range
$y=2-3$ to an average $dN/dy \approx 12$.
Baryon conservation in the reactions can be exploited to set limits
on the relative rapidity loss and the energy per baryon at RHIC. 
This is illustrated in Fig.~\ref{fig:dyvs_energy}, which in the insert shows two possible distributions whose
integral areas correspond to the number of baryons present in the
overlap between the colliding nuclei. From such distributions one can  deduce a set of upper 
and lower limits for the rapidity loss at RHIC. Since not all baryons are measured with the bulk of these in the rapidity interval
3-5.4 these assumptions have to be made for these.
The limits shown in the figure includes
estimates of these effects~\cite{BRAHMS:2004netproton} for different extrapolations.
The conclusion is that the absolute rapidity loss at RHIC $(\delta y =2.05
\pm 0.17)$ is slightly larger than at SPS. The value is
close to expectations from extrapolations of pA data at lower
energies ~\cite{BuszaGoldhaber,BuszaLedoux}. In fact the relative
rapidity loss is significantly reduced as compared to an extrapolation of the low energy
systematic~\cite{FVOH}.

Also from these distribution one can estimate the average energy loss of the
colliding nuclei.
We  find this to be about $73\pm 6~{\rm GeV}$ per
nucleon, but still with a significant uncertainty if the extreme limits
are assumed. The same limits on the relative rapidity loss gives a range of energy loss per baryon of
$ 47~{\rm GeV} < E < 85 ~{\rm GeV}$. 

\section{Particle Ratios}

Ratios of baryon to anti-baryons also give information on baryon transport
albeit indirectly since it depends on both transport and baryon pair-production. 
In particular the fraction $p/\bar{p} -1 $ is 
the relative fraction of transported vs. produced protons. At RHIC several experiments have measured properties
of the particle ratios. It is found that the centrality dependence of \pbarp is weak 
\cite{phenix:2004hadrons200,phenix:2004hadrons130,brahms:2005y01},
the \pT-dependence up to several GeV/c is flat, and the ratios of anti-neutrons to neutrons was deduced
from measurements of $\bar{d}/d$ \cite{phenix:2004dbard} and found to be consistent with that of protons.
In the following I will discuss  results on \pbarp from PHOBOS on collision geometry  and centrality dependence  
and from BRAHMS on rapidity dependence both being compared to models.

\begin{figure}[!ht]
\begin{center}
\resizebox{0.50\textwidth}{!}
           {\includegraphics{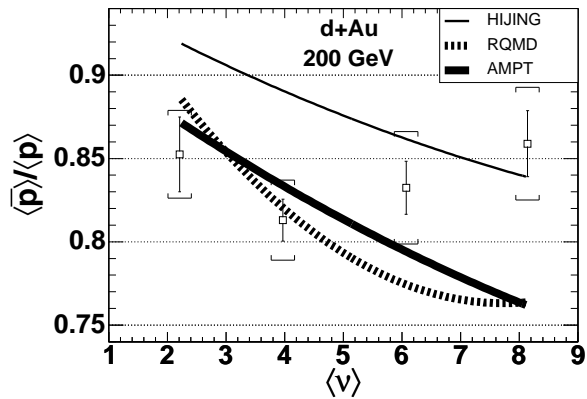}}
\end{center}
\caption{\label{fig:phobos_dau} Ratios of anti-proton to proton as function of centrality in d-Au collisions at \srhic. 
The figure is from the Phobos Collaboration \cite{phobos:dau_ratios}
}

\end{figure}

The particle ratios near mid-rapidity has been measured in d+Au, Au+Au 
and pp collisions by the PHOBOS collaboration\cite{phobos:dau_ratios,phobos:ppratios}. 
The \pbarp ratios in d-Au are very close to that observed in pp collisions (\pbarp$\simeq 0.84$) 
while larger than what is seen in AuAu collisions (\pbarp$\simeq 0.76$ ).
This indicates that a larger fraction of baryon are transported to mid-rapidity 
in Au-Au collisions than in pp and d-Au collisions. 
This is consistent with the expectation of additional scattering in the heavy ion system.
The surprising observation is that when the d-Au system is studied versus centrality, or rather the mean number of
collisions $<$$\nu$$>$ estimated from a Glauber description is consistent with no dependence. 
This is shown in Fig.\ref{fig:phobos_dau} taken from their publication. 
This is in contrast to results from calculations of HIJING \cite{HIJING}, AMPT\cite{AMPT} and RQMD\cite{RQMD} model that all predict
a significant decrease of \pbarp with $<$$\nu$$>$. 
Such behavior arises naturally in the picture where multiple collisions cause increasing stopping and baryon transport, but is
apparently not born out by the data from RHIC.

\begin{figure}[!ht]
\begin{center}
\resizebox{0.55\textwidth}{!}
           {\includegraphics{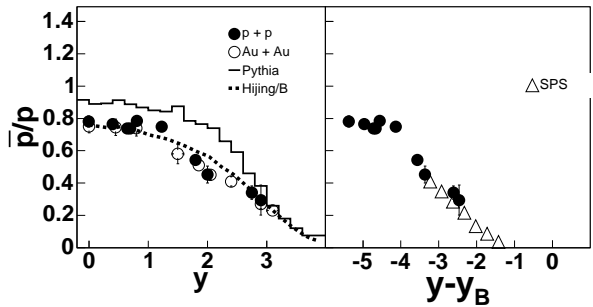}}
\end{center}
\caption{\label{fig:ratiosComparison}
 Left: \pbarp from p+p at \snn (solid) points compared with Au+Au \cite{BRAHMS:2002ratios} (open points), 
and predictions from PYTHIA \cite{Sjostrand:2000wi} (solid histogram) and HIJING/B \cite{Vance:1998vh} (thick dashed line). 
Right ratios shifted by $y_b$ compared with data from NA27 (triangles) at \snn =27.5 GeV \cite{Aguilar-Benitez:1991yy}.
}

\end{figure}

BRAHMS has recently presented measurements and analysis of $p/\bar{p}$ ratios 
in pp collisions at 200 GeV as function of rapidity \cite{BRAHMS:2005ppratios}.
Figure~\ref{fig:ratiosComparison} shows the resulting ratios of
antiparticle to particle yields as a function of rapidity (left
panel). For the ratios  there is a
clear midrapidity plateau and subsequent decrease with rapidity.
This $\bar{p}/p$ ratio would implies that at midrapidity 12\% of
the protons carry baryon number that has been transported from the
beam region at $y=5.3$.
It has been shown (see Ref.\cite{Fischer:2002qp}) that one may need to correct
for isospin effects before generalizing these results from $p+p$ to
hadron--hadron collisions.
At $y < 1.5$ the Au+Au ratios for the 20\% most central
collisions reported in~\cite{BRAHMS:2002ratios} are noticeably similar to
the present results. 
The kaon (not shown) and proton ratios remain consistent with the Au+Au results over
our entire acceptance range. This is surprising in view of the
different dynamics one might expect for the two systems. 
The ratio starts to decrease above $y=1.5$, indicating a transition from
the string breaking dominated regime at midrapidity to the
fragmentation region. 
Though results for pp and AuAu looks similar in term of rapidity dependence there is a difference
at mid-rapidity with the \pbarp values being lower in pp, as also shown by the PHOBOS results discussed above.

The right panel of Fig.~\ref{fig:ratiosComparison} shows the present
data and data from NA27 at \sqrts{27.5}~\cite{Aguilar-Benitez:1991yy}
(open triangles) shifted by the respective beam rapidities. Overlaying
the two datasets we observe the ratios to be independent of the
incident beam energy when viewed from the rest frame of one of the
protons in the area where our rapidity region overlaps with he other
experiment. This is consistent with the idea of limiting fragmentation
that has also been observed for charged hadrons in nucleus--nucleus
collisions~\cite{Bearden:2001qq,Deines-Jones:1999ap,PhobosFrag}. 
We also note a transition
in behavior at $y-y_{b}=-4$, indicative of a boundary between the
midrapidity and fragmentation regions.

To interpret these results further, we confront predictions from
theoretical models of hadron-hadron collisions with the data.  The
curves in the left panel of Figure~\ref{fig:ratiosComparison} compare
our results to the predictions of two such calculations, PYTHIA
Ver. 6.303~\cite{Sjostrand:2000wi} and HIJING/B~\cite{Vance:1998vh}, using the same
$p_{\rm{T}}$ range as the present analysis.  Both models give a good
description of the pion data and for kaons at midrapidity.
Also, PYTHIA clearly overestimates the $\bar{p}/p$ ratios.
This is a well--known problem since PYTHIA employs only quark--diquark
breaking of the initial protons.

The baryon junction scenario, incorporated as a model
prediction in the HIJING/B event generator~\cite{Vance:1998vh}, is shown as the dashed lines
in Fig.~\ref{fig:ratiosComparison}, exhibit a much better
agreement with the data both in terms of overall magnitude and the
width of the distribution. In Ref.~\cite{Capella:sx} the author shows
that baryon stopping in $p+p$ and Au+Au collisions at SPS and RHIC
energies can be described using the same parameters for the baryon
junction couplings, and predict that at RHIC the shapes of the
rapidity distributions for $p+p$ and Au+Au will be similar for
$|y| <  2$. The similarity of $\bar{p}/p$ in $p+p$ and $A+A$ up to
$|y| < 3$ supports this prediction.

\section{Model comparisons to Au-Au collisions at RHIC}

In the following section comparisons to models at RHIC energies will be made.
A large number of studies have been carried out both at lower energies 
and shows that the cascade mechanism via multiple collisions and resonances excitation is dominant at SIS and AGS energies.
Though, already at SPS energies the stopping neither by cascade model nor by string models.
In Ref.\cite{Vance:1998vh} the baryon junction mechanism was introduce to enhance stopping
over what a conventional string description would give and achieved a satisfactory description of the  NA49 net-proton data.
In this paper there is also a prediction for RHIC which in fact over estimates the actual later measurements 
of protons and $\Lambda s$.

\begin{figure}[!ht]
\begin{center}

 \begin{minipage}[t]{.45\textwidth}
    \includegraphics[width=\columnwidth]{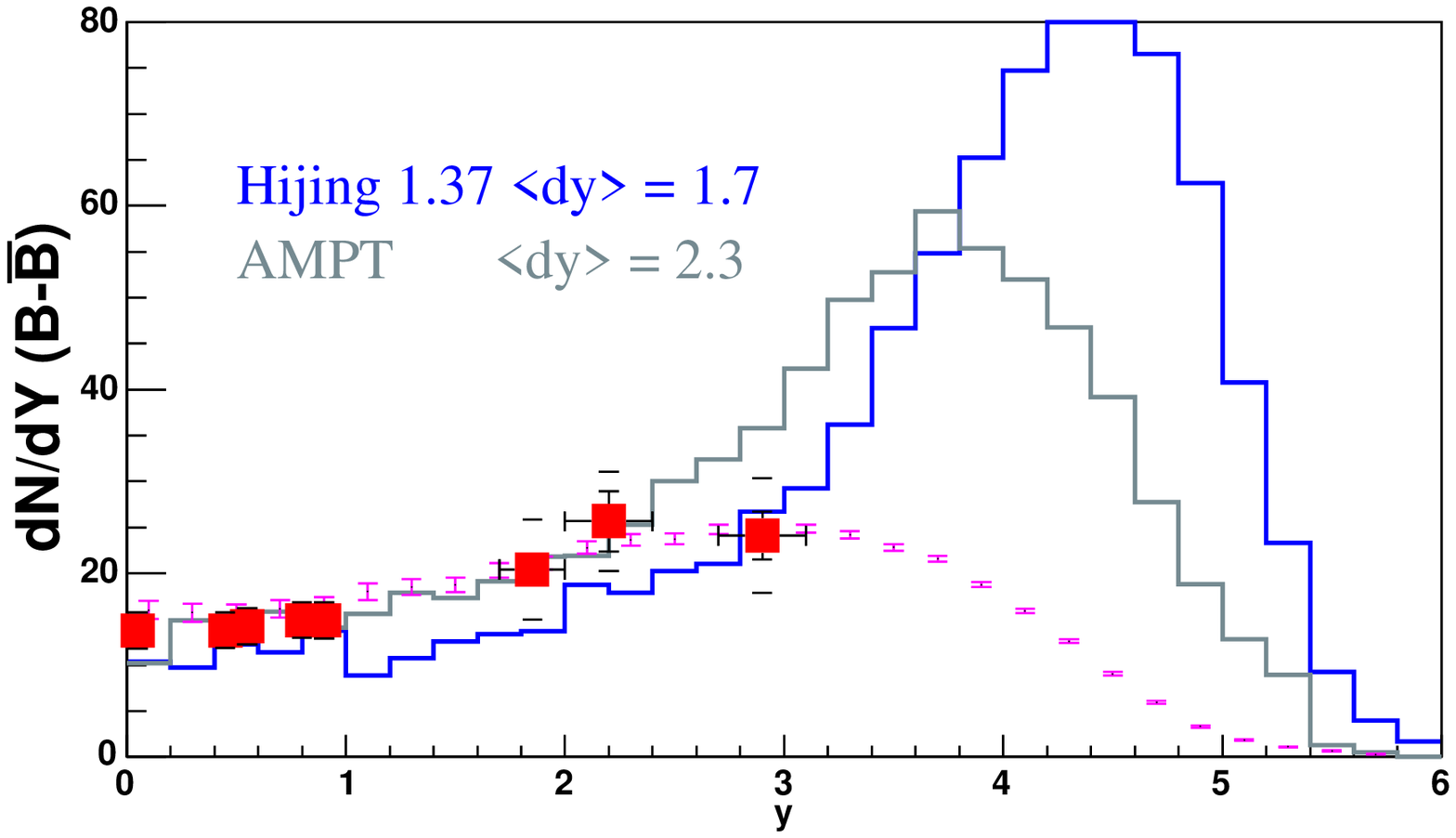}
  \end{minipage}
  \begin{minipage}[t]{.45\textwidth}
    \includegraphics[width=\columnwidth]{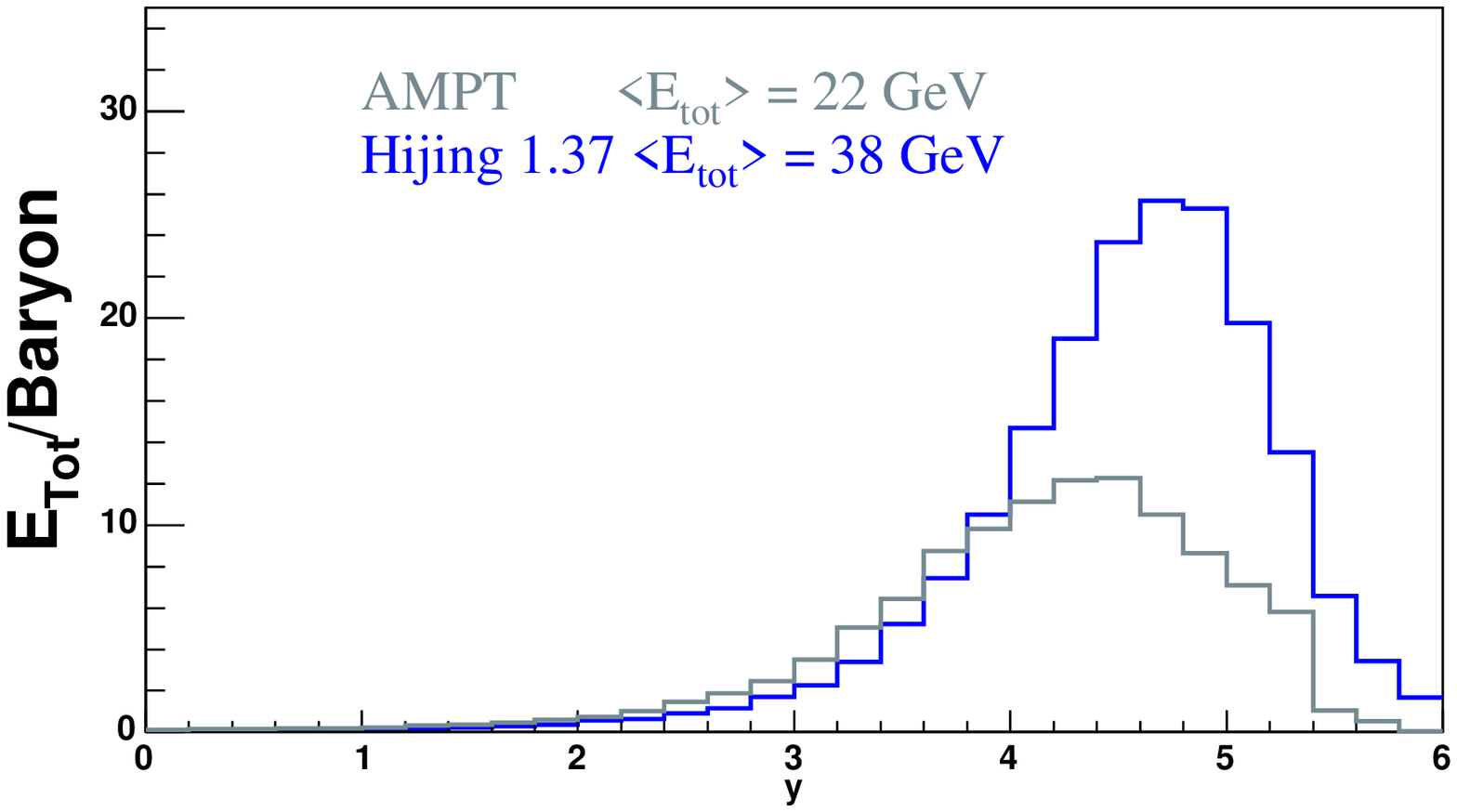}
  \end{minipage}

\end{center}
\caption{\label{fig:netbar_models}The left panel shows the rapidity density distributions compared to AMPT, 
HIJING and parton cascade model calculations. The right panels shows the rapidity distributions of energy per baryon.}

\end{figure}

In the left panel Fig.~\ref{fig:netbar_models} the net-baryons measured by BRAHMS are compared to 3 models.  
The HIJING\cite{HIJING} model (full drawn curve) where the main mechanism for baryon transport is $q-qq$ string breaking result 
in a net-baryon yield at mid-rapidity slightly lower than the data, and with a mean rapidity loss  of 1.7 being at the low end of the
allowed range. The AMPT model (light line)  on the other hand results in a much higher yield at mid-rapidity, still compatible with the data, 
but a mean rapidity loss at the upper end of the range. 
This model~\cite{AMPT} includes the socalled popcorn mechanism to describe baryon - antibaryon production with parameters adjusted
to described the NA49 data from SPS. 
The larger difference between the models are at the higher rapidities where the data are still lacking.
The models results in significant different energy per baryon in the final state of 38 and 22 GeV, respectively. 
On the figure is also shown the calculation of the parton cascade description~\cite{bass:2003baryons} (dashed points). 
This particular model should only be compared with data near mid-rapidity,
since the spectator baryons that are left within a few units of beam rapidity are not explicitely dealt with in this model.
In conclusion several models do in fact describe the measurements at RHIC energies, but the conventional string breaking
description does underestimate the baryon transport to mid-rapidity.

Recently the HIJING/B model was improved in Ref.\cite{ToporPop:2004} by taking into account intrinsic $k_T$ motion.
The main idea is to see if the baryon junction mechanism is also able to account for the enhanced baryon over meson 
ratios seen in AA collisions in the intermediate \pT-range of 1-4 GeV/c.
The model was also compared to the BRAHMS proton and net-proton distributions. 
The results are available in Ref.\cite{ToporPop:2004} as Figs. 4 and 6 and show an overall good description.
The authors also calculated the mean rapidity loss to be 1.75 similar to that of the pure HIJING calculation shown in Fig.\ref{fig:netbar_models}
and a mean energy per baryon of 40 GeV. This number are within the experimental values of $2.0 \pm 0.2$ albeit on the lower range;
as a result the mean energy is also considerable large than the value of 26~GeV quoted here
albeit as pointed out this experimental has an considerable uncertainty. 
The model thus does transport additional baryons to mid-rapidity being a candidate
for the correct description. 
Additionaly this transport is an important component in describing the enhancement of baryon over pions
observed in the intermediate \pT ~range of 1-5 GeV/c. 
As stated by the authors it fails in describing the transverse spectra of kaons and $\Lambda s$.
All in all there are indications that the baryon junction picture are important for baryon dynamics at RHIC. 

\section{Summary}

At the highest energies so far available for heavy ion collisions, namely \srhic ~at RHIC the mean rapidity loss in
AA collisions seems to have reach an approximate saturated value of $\approx 2$ units. 
In contract the measurements of net-proton distributions so far do not constrain the mean energy deposited in the collision as well,
i.e. the energy that is available for particle production , longitudinal and transverse motion. 
Evaluation of the data, as well as comparison to models indicate that a likely range of energies are $25-37$ GeV/ net-baryon
leaving $63-75\%$ for production.
The analysis of the distributions as well as the $p/\bar{p}$ values $< 1$ in both pp and AA seem to require additional baryon 
transport mechanism over quark diquark breaking.
The data still leaves open whether this is caused by the baryon junction mechanism, if it can be explained by other of the
proposed mechanisms. 
Such mechanisms as these will in general not decrease the energy per baryon, since only baryon number 
is transported to mid-rapidity while the energy associated will reside at large rapidities in forward going mesons. 
Thus the direct connection between stopping of energy and rapidity loss of net-baryons is broken at the higher energies.

The outlook to have some of these questions clarified is encouraging. Not only are there more data to come from RHIC at both
\sqrts{62.4} and \srhic, but also the upcoming ALICE experiment at LHC will have a distinct possibility to measure
the baryon asymmetry in pp and AA collisions with a rapidity loss of up to$ \sim 8-9.6$ units of rapidity which can help to 
disentangle the transport processes with different energy dependences.

\section{Acknowledgments}

This work was supported by the Division of Nuclear Physics of
the of the U.S. Department of Energy under contract DE-AC02-98-CH10886.
I also appreciate interesting discussions with S.Bass, Nu Xu, and H.Stocker on this subject, 
M.Murray and J.H.Lee for carefull reading of the manuscript, and  will like to thank the organizers for an interesting conference in Kolkata.

\end{document}